**Energy Transfer to a Stable Donor Suppresses Degradation in Organic Solar Cells**

*Andreas Weu, Rhea Kumar, Julian F. Butscher, Vincent Lami, Fabian Paulus, Artem A. Bakulin and Yana Vaynzof\**

Andreas Weu, Julian F. Butscher, Vincent Lami, Dr. Fabian Paulus, Prof. Yana Vaynzof

Kirchhoff Institute for Physics and the Centre for Advanced Materials, Im Neuenheimer Feld 227, 69120 Heidelberg, Germany

Email: vaynzof@uni-heidelberg.de

Rhea Kumar, Dr. Artem A. Bakulin

Centre for Plastic Electronics, Department of Chemistry, Imperial College London, London SW7 2AZ, United Kingdom

Abstract:

Despite many advances towards improving the stability of organic photovoltaic devices, environmental degradation under ambient conditions remains a challenging obstacle for future application. Particularly conventional systems employing fullerene derivatives are prone to oxidise under illumination, limiting their applicability. Herein, we report on the environmental stability of the small molecule donor DRCN5T together with the fullerene acceptor $PC_{70}BM$. We find that this system exhibits exceptional device stability, mainly due to almost constant short-circuit current. By employing ultrafast femtosecond transient absorption spectroscopy we attribute this remarkable stability to two separate mechanisms: 1) DRCN5T exhibits high intrinsic resistance towards external factors, showing no signs of deterioration. 2) The highly sensitive $PC_{70}BM$ is stabilised against degradation by the presence of DRCN5T through ultrafast long-range energy transfer to the donor, rapidly quenching the fullerene excited states which are otherwise precursors for chemical oxidation. We propose that this photoprotective mechanism be utilised to improve the device stability of other systems, including non-fullerene acceptors and ternary blends.

## 1. Introduction

Organic photovoltaics (OPVs) have seen rapid performance improvements in the last five years, now achieving power conversion efficiencies (PCEs) of 16%.[1,2] While most OPV devices incorporate polymers as donor materials, there is an increasing interest in applying solution-processable small molecules as both the donor and acceptor components of the bulk heterojunction (BHJ) active layer. This is motivated by their easy purification, small batch-to-batch variation and improved stability relative to polymer materials.[3–7] The application of small molecules as non-fullerene acceptors (NFA) is already common, the emergence of which resulted in an unexpected surge in device performance.[8–14] Fewer examples exist of small molecules applied as donor materials,[15,16] but efficiencies surpassing 10% have been demonstrated.[17,18]

With increasing performances offering OPVs as a promising candidate for industrial application, the remaining challenge of low operational stability precludes their large-scale integration into the PV market. The degradation mechanisms of polymer:fullerene-based solar cells have been a focus of extensive study in the last decade and are now reasonably well understood. Under illumination, polymer solar cells typically show a strong initial decrease in device performance ('burn-in'), caused by formation of defects in the bandgap.[19–22] In addition, exposure to oxygen leads to p-type doping in some polymer systems, which causes increased bimolecular recombination or formation of traps, which in turn impede charge generation and transport.[23–26] The combined influence of oxygen and light induces irreversible photo-oxidation in many polymers through reactions involving singlet oxygen or radical superoxide, causing chemical lesion and deactivation of the chromophores.[27–29] Fullerene acceptors have been reported to dimerise or degrade upon exposure to oxygen and light, which limits device performance due to trap formation or a reduction in electron mobility.[30–33] Morphological changes induced by elevated temperatures under operating conditions provide another degradation pathway. The most crucial feature required for efficient OPVs is a finely mixed, bicontinuous BHJ comprising donor and acceptor domains of sizes on the order of the exciton diffusion length (~10 nm).[34–37] It has been shown that exposure to light or high temperatures causes demixing of the phases thus reducing propensity for charge separation, with this demixing often accelerated by the presence of processing additives like DIO. [19,38–41]

In contrast to polymer:fullerene systems, the stability of small molecule (SM) OPVs has attracted little attention so far. Early work investigated the influence of water and oxygen in zinc phthalocyanine-based solar cells, but identified either the low work function electrodes and transport layers in the standard architecture or the $C_{60}$ acceptor as the main factors for device failure.[42–44] While the poor stability of standard architecture devices in air are known,[42,45–48] inverted architecture SM OPVs have been reported to be relatively stable, but no justification of this difference or underlying degradation mechanism has been reported.[49]

The stability of 'new-generation' small molecule donors is even less well studied. One particularly interesting small molecule donor 2,2'-[(3,3''',3'''',4'-tetraoctyl[2,2':5',2'':5'',2''':5''', 2''''-quinquethiophene]-5,5''''-diyl)bis[(Z)-methylidyne(3-ethyl-4-oxo-5,2-thiazolidinediyliden

e)]]bis-propanedinitrile (DRCN5T) was first designed and synthesised by Y. Chen and co-workers in 2015.[50] This donor SM has been shown to result in impressive performance with the fullerene acceptor [6,6]-Phenyl-C70-butyric acid methyl ester (PC$_{70}$BM) as well as several NFAs.[51–53] The stability of various standard architecture SM OPVs (including DRCN5T:PC$_{70}$BM) has been reported by Cheacharoen *et al.* who showed that heating the films, either on a hotplate or during exposure to visible light, leads to significant burn-in within 100 hours, attributed to morphological changes in the nanostructure of the blend.[54] Min *et al.* investigated the stability of encapsulated standard architecture DRCN5T:PC$_{70}$BM devices with a focus on the role of the active layer microstructure on the degradation dynamics.[55] However, no clear understanding of the degradation mechanisms and their effect on the photophysics of the blend have been reported to date.

In this work, we demonstrate that the representative SM system DRCN5T:PC$_{70}$BM exhibits superior stability relative to many other polymer:PC$_{70}$BM OPV systems. We show that this is due to a photoprotective mechanism effected by the donor molecule that suppresses the degradation of the fullerene acceptor. This photoprotective effect is enabled by two factors: (1) the intrinsically high photostability of the donor molecule, even in the presence of oxygen and (2) efficient and ultrafast energy transfer from the PC$_{70}$BM acceptor to the donor SM, which effectively quenches the high energy excited state on the fullerene, preventing its photo-oxidation. It is noteworthy that recent reports have identified fullerene degradation as a major limitation to the stability of OPV devices,[27,56,57] highlighting the importance of developing mitigation strategies such as the photoprotection realised herein. By applying ultrafast femtosecond (fs) transient absorption spectroscopy (TA), we examine the processes governing charge generation in both pristine and degraded solar cells and demonstrate how efficient charge generation is maintained upon exposure to oxygen and light. Finally, we discuss how this photoprotective mechanism can be exploited in future small molecule OPV devices.

## 2. Results

The chemical structures of DRCN5T and PC$_{70}$BM, along with the inverted architecture structure of the investigated solar cells are illustrated in **Figure 1**. This structure with high work function contacts was chosen to minimise degradation effects by the electrodes and allowed us to focus on the processes affecting the active layer.[26,49] In this architecture, caesium-doped zinc-oxide (ZnO:Cs) served as electron transport/hole blocking layer[58,59] and molybdenum-oxide (MoO$_x$) as hole transport/electron blocking layer; indium tin oxide (ITO) and silver acted as the contacts. In this first report of the application of DRCN5T:PC$_{70}$BM in an inverted device architecture, we were able to achieve a maximum PCE of 8.2 %, competitive with previously reported standard architecture devices.[50,55] **Figure 1**c shows the energy levels of the active components with respect to vacuum level, which have been measured by ultraviolet photoelectron spectroscopy (UPS) and UV/Vis absorption spectroscopy. Arrows indicate electron transfer from donor to acceptor and possible energy transfer from acceptor to donor, facilitated by overlapping absorption and emission of DRCN5T and PC$_{70}$BM (Figure S1).

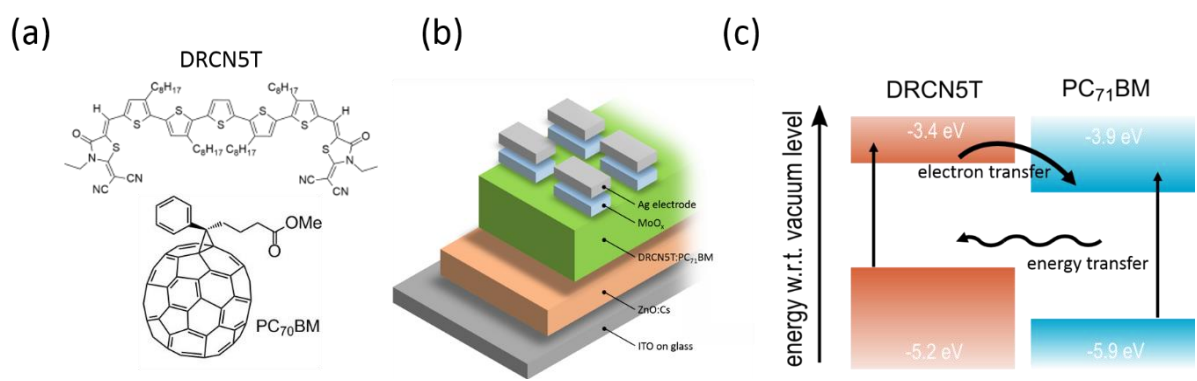

Figure 1. (a) Chemical structures of the small molecule DRCN5T together with the fullerene PC$_{70}$BM. (b) The inverted architecture solar cell comprising the active layer sandwiched between the ZnO electron transporting layer and MoO$_x$ hole transporting layer. (c) Energy level

diagram of the active layer with arrows indicating photon absorption, charge transfer at the interface and energy transfer between the materials.

To investigate the stability of this SM system upon exposure to oxygen and light, the performances of 24 DRCN5T:PC$_{70}$BM solar cells were monitored for 22 h under continuous illumination conditions of 1 Sun and a controlled atmosphere of 20% O$_2$ in a bespoke environmental setup.[26]

**Figure 2**a shows the effect of the aforementioned degradation conditions on the external quantum efficiency (EQE) spectra of the devices, acquired *in situ*. The response within the spectral range of 450 to 800 nm, arising predominantly from the contribution of the donor, decreases slightly due to a reduction in DRCN5T absorbance (see Figure S1). This decrease in optical absorption can arise due to photobleaching mechanisms like oxygen-induced radical formation or light-induced chemical modification of the molecular structure.[60–62] The decrease in EQE is more noticeable at wavelengths below 410 nm, where PC$_{70}$BM also contributes to current generation. Here, a clear reduction in the EQE is observed which cannot be correlated to the reduction of optical absorption (Figure S1). However, the overall decrease in quantum efficiency is unusually small when compared to other OPVs employing fullerene acceptors. For comparison, we show the EQE evolution of a poly [(2,6-(4,8-bis(5-(2-ethylhexyl-3-chloro)thiophen-2-yl)-benzo] [1,2-b:4,5-b' ]dithiophene))-alt-(5,5-(1',3'-di-2-thienyl-5',7'-bis(2-ethylhexyl)benzo [1',2'-c:4',5'-c']dithiophene-4,8-dione) (PBDB-T-2Cl):PC$_{70}$BM solar cells degraded under identical conditions. For this system, a strong reduction of the quantum efficiency is observed, as is typical for polymer:fullerene OPVs. To highlight the superior stability of DRCN5T:PC$_{70}$BM, the decrease of short-circuit current (J$_{SC}$) over time is compared to other high efficiency polymer:fullerene systems (**Figure 2**b). The benchmark materials poly[(5,6-difluoro-2,1,3-benzothiadiazol-4,7-diyl)-alt-(3,3'''-di(2-octyldodecyl)-2,2',5',2'',5'',2'''-quaterthiophen-5,5''''-diyl)] (PffBT4T-2OD), poly [4,8-bis]

[(2-ethylhexyl)oxy ]benzo [1,2-b:4,5-b' ]dithiophene-2,6-diyl [3-fluoro-2-] [(2-ethylhexyl)carbonyl ]thieno [3,4-b ]thiophenediyl [1,2 ] (PTB-7), and PBDB-T-2Cl, each with $PC_{70}BM$ as the acceptor, have been chosen as a comparison. While all polymer systems, particularly PffBT4T-2OD, show significant burn-in through loss of $J_{SC}$, the $J_{SC}$ of DRCN5T:$PC_{70}BM$ maintains over 90 % after 18 hours of degradation. To highlight the unusually stable current generation in DRCN5T:$PC_{70}BM$, we compare the results to a reference measurement taken in nitrogen under dark conditions (Figure S2), in which no reduction in $J_{SC}$ was observed.

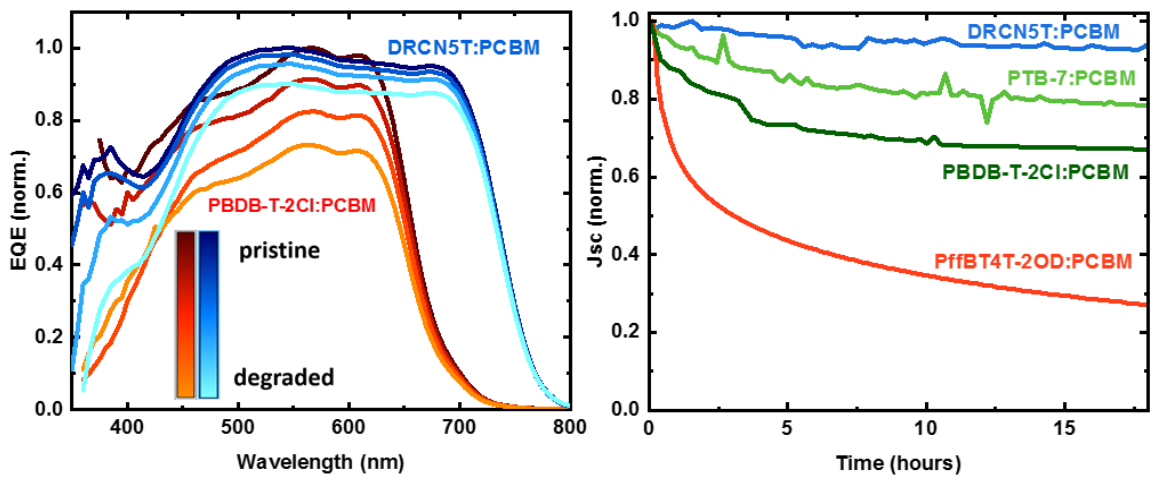

Figure 2. (a) Evolution of EQE spectra for DRCN5T:$PC_{70}BM$ and PBDB-T-2Cl:$PC_{70}BM$ solar cells upon exposure to light and oxygen for 22 h. (b) Reduction in $J_{SC}$ in DRCN5T:$PC_{70}BM$ as compared to three analogous polymer:fullerene OPV systems.

Both the spectral position and the low extent of the EQE reduction indicate that degradation in this system is effectively suppressed and mostly depletes photocurrent generated due to absorption by the $PC_{70}BM$ molecules. Femtosecond transient absorption spectroscopy was therefore carried out in order to investigate the effect of degradation on the exciton and charge carrier dynamics of the system.

**Figure 3** shows the kinetics of neat films of DRCN5T and PC$_{70}$BM under the previously described degradation conditions; corresponding spectra can be found in Figure S3. A singular value decomposition (SVD) analysis was performed on the data which yielded a single dominant component for the DRCN5T film, assigned to donor excitons. Normalised kinetics of the charge pairs are plotted in **Figure 3**a for a DRCN5T film during 9 hours of continuous white light illumination in dry air. After excitation with a 700 nm pump pulse, excitons are created within the time resolution of the setup (100 fs) and recombine within 300 ps. No effect of degradation is observed in the exciton dynamics apart from ~ 20 % loss of absorption due to photobleaching (see Figure S3), demonstrating the intrinsically high stability of DRCN5T.

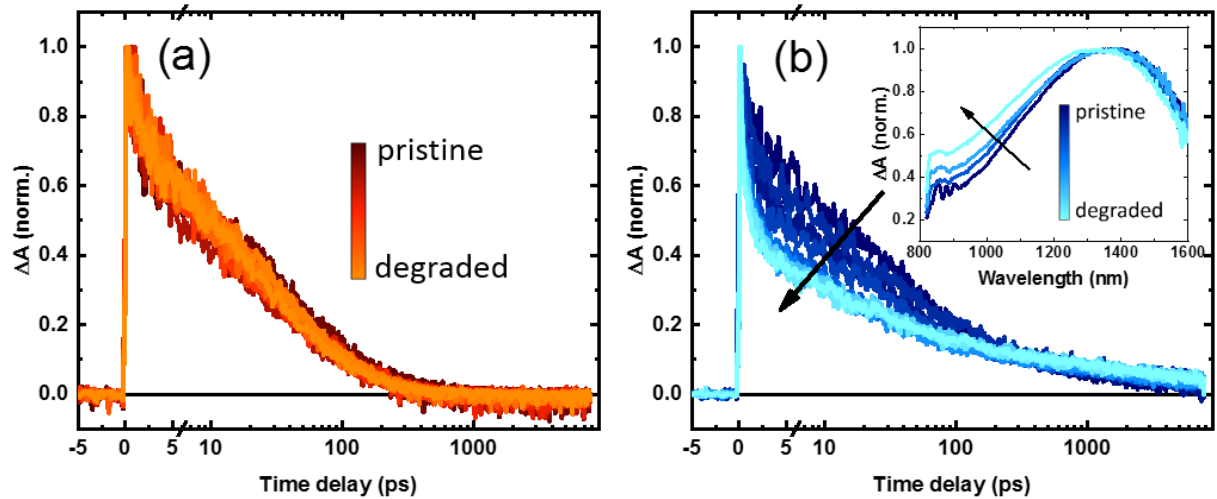

Figure 3. Transient absorption data for the neat materials in the studied blend. (a) Kinetics of DRCN5T excitons (excited at 700 nm with 150 nJ, probed at 1200 nm) do not change upon degradation. (b) Kinetics of PC$_{70}$BM excitons (excited at 350 nm with 150 nJ, probed at 1350 nm) show greater early quenching of excitons with increased degradation. Inset: spectral change indicating the formation of trapped species.

In the neat PC$_{70}$BM film, excitons were created by a 350 nm pump pulse. Prior to degradation, the excitons decay with a lifetime of 50 ps. However, continuous exposure to oxygen and light for 7 hours causes a notable reduction of exciton lifetime to 7 ps. After 300 ps this trend inverts as a marginally larger exciton amplitude with increasing degradation is observed. We speculate that the initial reduction of the signal followed by an increase at later times corresponds to the formation of trap states as a consequence of degradation, efficiently quenching the excitons and creating long-lived, trapped charges. Accordingly, the spectrum between 800 nm and 1250 nm changes gradually during exposure to oxygen and light (see inset Figure 3b), indicating the formation of a second species, likely to be trapped carriers.

Having investigated the stability of the constituent materials, we turn our attention to the degradation of the OPV blend system. **Figure 4** shows the dynamics of the system after selectively exciting the donor molecule with 700 nm pump pulses. Two distinct contributions to the blend spectrum, assigned to DRCN5T excitons and free charge carriers, were extracted by global analysis (GA) and are shown in Figure 4. Figure 4a depicts the spectrum of the DRCN5T exciton created directly by the pump pulse, visibly unaffected by exposure to light and oxygen. The kinetics shown in Figure 4b confirm this stability, with approximately consistent exciton decay lifetimes which vary slightly but without correlation to degradation (see Figure S6). The lifetime of donor excitons in the blend is intuitively shorter than that in neat DRCN5T due to quenching by charge transfer at the donor:acceptor interface. The kinetics associated with the second component reveal that free charge carriers form on a 10 ps timescale, concomitantly with the decay of excitons. The charges are characterised by a long-lived signal (> 8 ns), suggesting that recombination is primarily of a bimolecular nature. The spectrum of charges formed in the blend remains unchanged during degradation (Figure 4c), with lifetimes of charge generation and decay processes showing no clear trends during the degradation experiments (see Figure S6). Taken together, these observations suggest that the processes of exciton formation, diffusion to an interface and subsequent charge separation by electron transfer to PC$_{70}$BM are unaffected by exposure to oxygen and to 1 Sun illumination throughout the duration of the experiment. In contrast to the neat materials, no change in absorption is observed (see Figure S4) suggesting a stabilising influence of the blend environment on the excitons, most likely due to fast quenching of high energy excitations on the donor by the presence of the acceptor.[33,63]

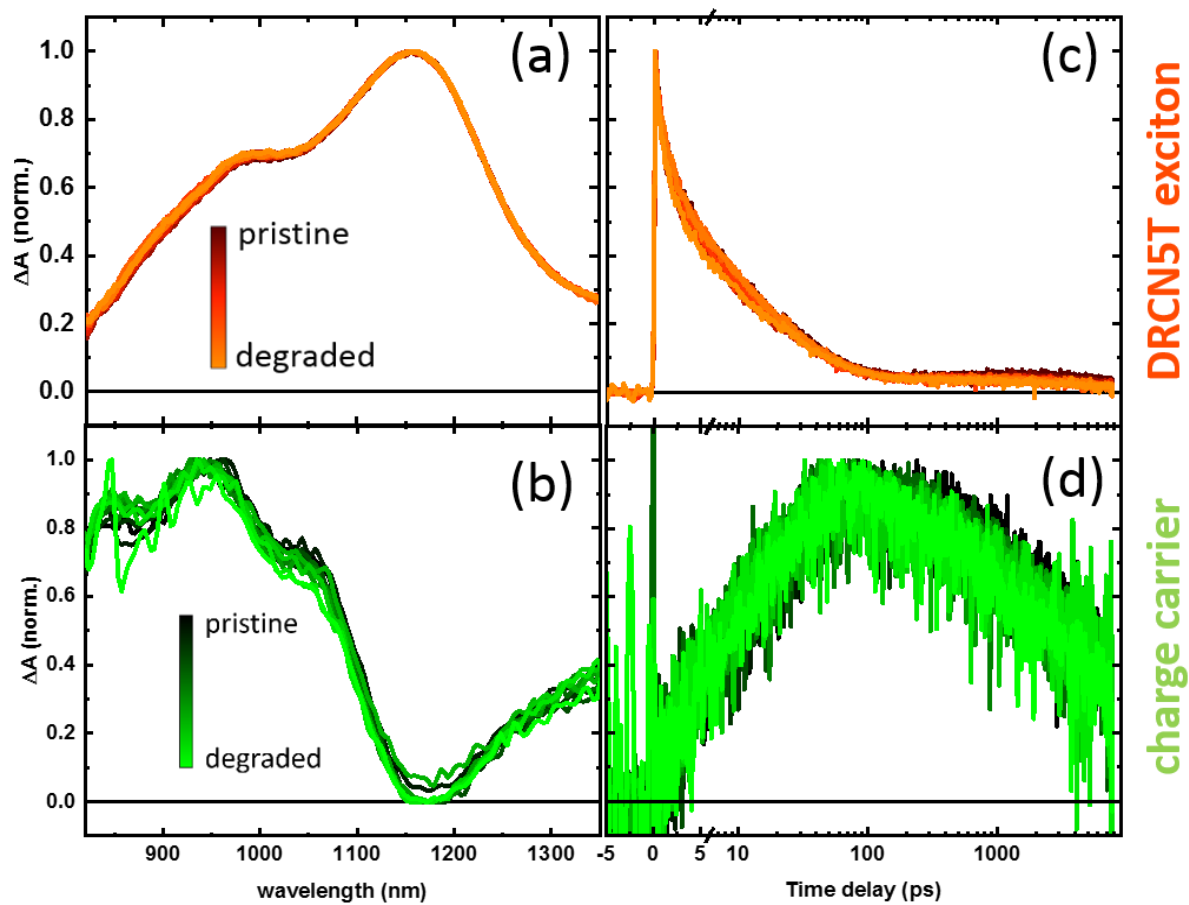

Figure 4. Spectra (left) and kinetics (right) of GA-derived components of the transient absorption spectrum of the DRCN5T:PC$_{71}$BM blend after 700 nm excitation (150 nJ) at selected stages of the degradation process. (a) DRCN5T exciton and (b) charge spectra show no signs of degradation. (c) Exciton and (d) charge kinetics are unchanged by degradation over 5 hours in dry air and a constant illumination of 1 Sun.

In order to investigate the role of the fullerene on the photophysical behaviour of the blend, TA of DRCN5T:PC$_{70}$BM was carried out using a 350 nm pump pulse for selective excitation of the acceptor. We note that a control experiment in neat DRCN5T (Figure S5) shows only a minor response of the donor to the 350 nm pulse. Figure 5 shows the normalised decay in amplitude of the resultant spectra at the selected wavelength of 1300 nm, representing the maximum absorption of the PC$_{70}$BM exciton, for neat PC$_{70}$BM (left) and for DRCN5T:PC$_{70}$BM (right). The effect of degradation by 1 Sun illumination and oxygen exposure is striking for the neat acceptor film, while the blend kinetics reveal an evident stability to such degradation in comparison.

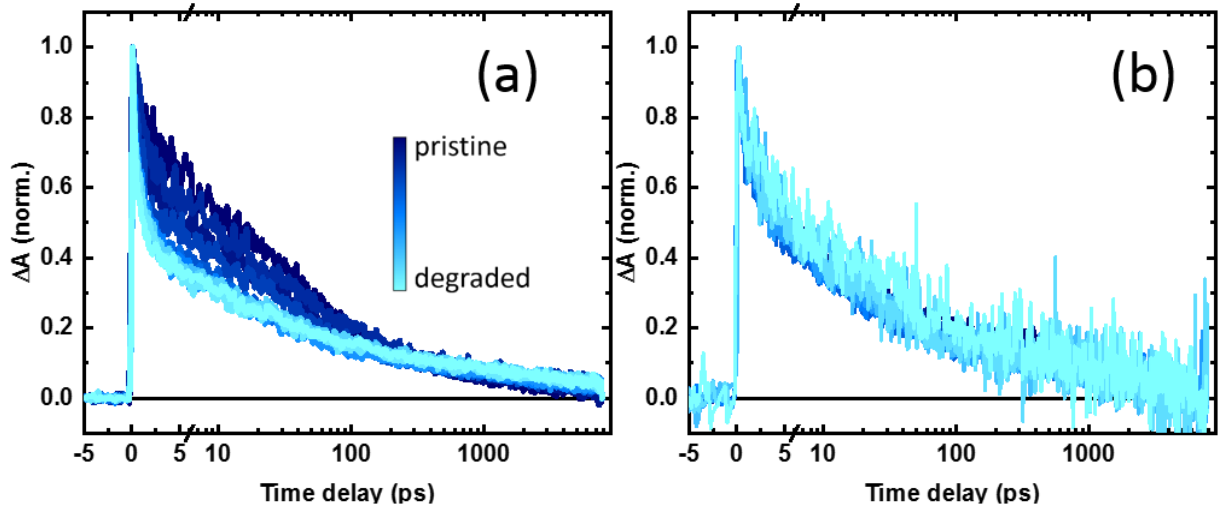

Figure 5. Normalised decay in amplitude of (a) neat PC$_{70}$BM and (b) DRCN5T:PC$_{70}$BM blend spectra excited with a 350 nm pump pulse (150 nJ) at the selected wavelength of 1300 nm at increasing extents of degradation.

As DRCN5T has proven stable both in pristine and blended environments, our results raise the question of the mechanism actively preventing the degradation of PC$_{70}$BM when blended with DRCN5T. In order to elucidate the origin of this remarkable enhancement in stability of the blend over its components, transient absorption spectra were also recorded following a 250 nm excitation. At this wavelength, a far larger fraction of PC$_{70}$BM is excited relative to DRCN5T. We note that the acceptor is not exclusively excited at this wavelength, but the absorption of the donor is minimal in comparison. The three components resulting from spectral reconstruction with GA are shown in **Figure 6**. As demonstrated in the normalised kinetics, only the PC$_{70}$BM exciton is generated by pump excitation. In contrast, the formation of the

DRCN5T exciton is not initiated by the pump pulse, but instead occurs concurrently with PC$_{70}$BM exciton decay, indicating the manifestation of efficient energy transfer from the acceptor to the donor on the sub-picosecond timescale. Based on these observations we propose the following model (illustrated in Figure 6c) to rationalise the stabilising nature of the blend. Excitation of PC$_{70}$BM gives rise to an ultrafast and long-range energy transfer to DRCN5T, enabled by spectral overlap of PC$_{70}$BM emission and DRCN5T absorption (Figure S1). The resulting donor exciton diffuses to an interface where it undergoes a subsequent dissociation of the bound state into long-lived charge carriers after 100 ps. Since DRCN5T is intrinsically stable, the energy transfer does not lead to its degradation and rather facilitates efficient photoprotection of the fullerene.

The observations show that exciton formation and diffusion, electron transfer at the interfaces, and charge separation in DRCN5T:PC$_{70}$BM devices is unaffected by the presence of oxygen or light. However, the observation of a stable short-circuit current suggests that also the transport of charges to the electrodes remains largely unaffected, suggesting no significant changes to the active layer microstructure. To confirm this, we employed 2D-XRD measurements to determine the microstructure and crystallinity of the blend before and after degradation (see Figure S7a and b). The results show no substantial changes in the microstructure due to the degradation in oxygen and light for 20 hours, from which we conclude, that that electron and hole transport through the active layer remains stable throughout the entire experiment and oxygen exposure. . Furthermore, transient photocurrent (TPC) kinetics were measured to analyse the extraction dynamics at the interlayers/electrodes (Figure S7c). We find no correlation between extraction time (typically influenced by the presence of deep traps) and degradation, indicating that charge extraction across the interfaces to the transport layers and electrodes is predominantly unaffected by oxygen or light.

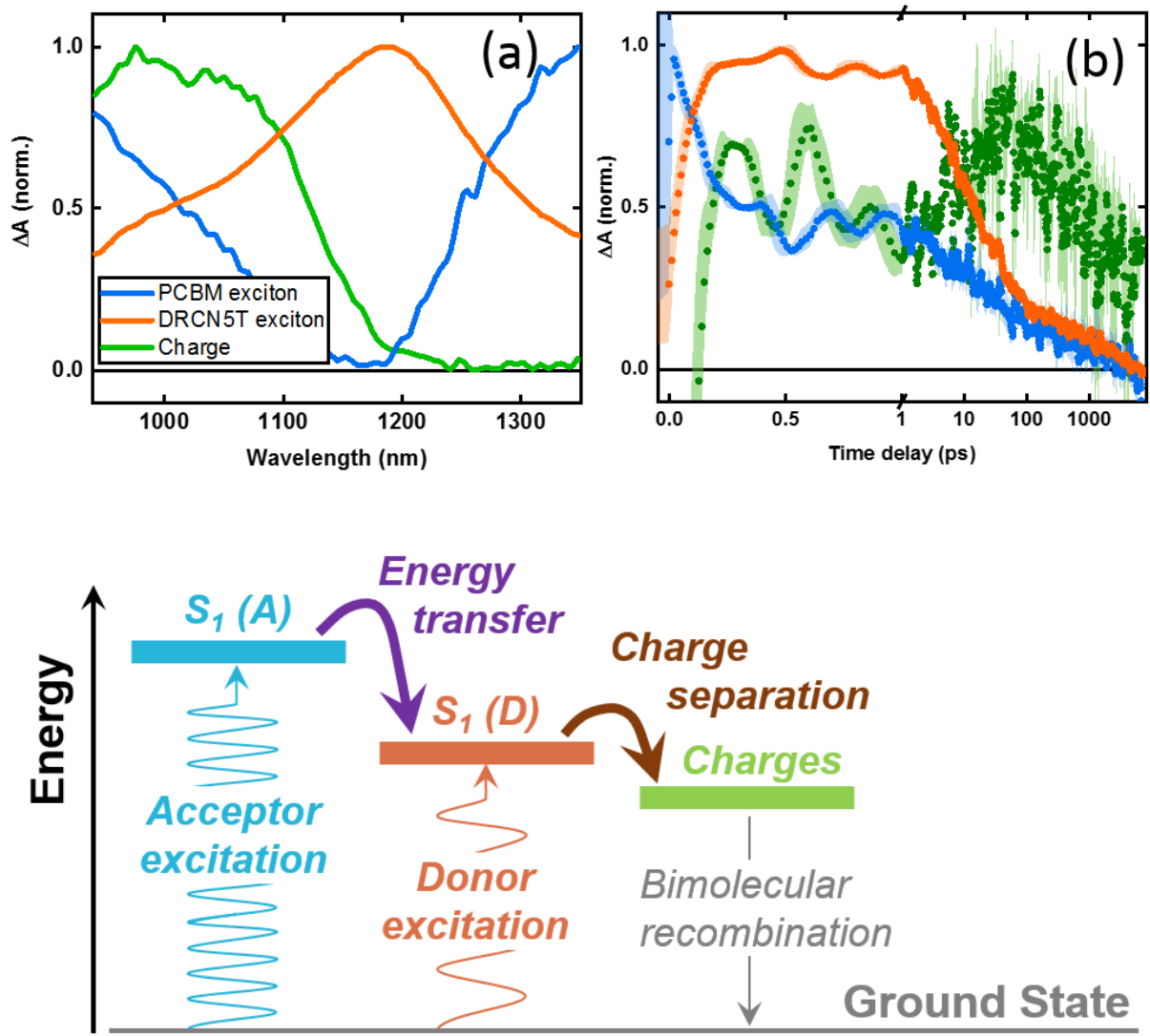

Figure 6. Normalised spectra (a) and kinetics (b) for the three contributions to the TA spectrum of DRCN5T:PC$_{70}$BM after 250 nm excitation (150 nJ): acceptor exciton (blue), donor exciton (red), and free charges (green). Kinetic traces are plotted as moving averages with associated errors shown. (c) Proposed model of energy transfer from PC$_{70}$BM to DRCN5T prior to charge transfer.

## 3. Discussion

The presence of an efficient, ultrafast energy transfer has fascinating implications for the photophysics of this material system. In this case, the excitation of fullerene molecules does not result in a hole transfer to the donor, since charge transfer occurs on a timescale of ~10 ps, while energy transfer is far faster, on a sub-ps timescale. It is important to note that this does not prevent the absorption by the acceptor from contributing to photocurrent. Instead photocurrent is generated through electron transfer from the donor that follows the energy transfer process. By designing novel photovoltaic systems with complementary absorption to facilitate such energy transfer, one can maintain the contributions of both active layer components, while eliminating the limitations originating from the instability of fullerene or other wide-bandgap acceptors. Energy transfer also makes material insensitive to the rate of hole transfer and allows to focus the material design exclusively on the optimization of electron transfer. Additionally, long-range energy transfer can assist in current generation by directed exciton transport to a donor:acceptor interface in contrast to range-limited, random exciton diffusion.[64]

The combination of the energy transfer from the acceptor with an ultra-stable donor, like in DRCN5T:PC$_{70}$BM results in both the charge generation and transport/extraction mechanisms remaining largely unaffected upon prolonged exposure to oxygen and light. Consequently, the J$_{SC}$ in such a material system remains very stable over an extended period of time, in stark contrast to conventional OPV systems. We believe that the observed photoprotective effect of energy transfer from acceptor to donor can be further exploited in other systems and should influence design rules towards more stable devices. Current research focuses exclusively on the development of novel photovoltaic systems with efficient charge generation, in which the process of charge transfer places a key role. Our work demonstrates that the selection of active components must be done with consideration also of complementary absorption and emission spectra, as long-range energy transfer can stabilise the blend *via* ultrafast quenching of the excited state which would otherwise lead to rapid degradation of the material. Such a photoprotective effect essentially relaxes the stability requirement to only one of the blend components, while the other, higher bandgap, material can be less stable and yet remain protected. This effect could be particularly useful for NFAs with a low electron affinity, which enable high $V_{OC}$ but simultaneously facilitate formation of radical oxygen species by electron transfer.[65–67] We further propose that the stabilising character of DRCN5T or similarly stable materials can be further capitalised in other systems employing unstable acceptors, for example by incorporating it as a third component in a ternary blend.[68] A preliminary experiment in PBDB-T-2Cl:PC$_{70}$BM solar cells showed that the stability of $J_{SC}$ can be increased by 15 % upon the addition of 20 % DRCN5T to the blend (see Figure S8).

## 4. Conclusion

In this work, we conclude that the SM:fullerene system DRCN5T:PC$_{70}$BM is remarkably stable under continuous illumination in ambient conditions, evidenced by an unprecedentedly consistent $J_{SC}$. We rationalise this observation with two arguments: 1) DRCN5T itself is photophysically very stable and shows no signs of degradation aside from a minor photobleaching effect. 2) The inherent instability of PC$_{70}$BM is overcome by ultrafast energy transfer from the fullerene to the small molecule preventing the otherwise detrimental effect of high energy excitations on the acceptor. These results highlight the superior stability of small molecule OPV materials, which should influence future strategy in the pursuit of efficient and stable material systems. Furthermore, consideration of complementary donor and acceptor molecules for design of blends that are specifically capable of ultrafast, long-range energy transfer can greatly improve the stability of the entire system. Finally, incorporation of DRCN5T into a ternary blend consisting of high-bandgap materials could lead much more stable device performance.

## 5. Experimental Section

**Materials**

DRCN5T and PC$_{70}$BM were purchased from Ossila and Solenne. Zinc acetate dihydrate as a precursor to ZnO was bought from Sigma Aldrich. All the materials were used as received without further purification.

**Fabrication OSCs**

The investigated organic solar cells were fabricated on pre-cut glass substrates, patterned with ITO as bottom electrode (PsiOTec Ltd., UK). After cleaning by subsequent sonication in acetone and isopropanol the substrates were transferred to a plasma cleaner for oxygen treatment. For inverted architecture devices, a thin ZnO layer (~50 nm) was spin-coated from solution and annealed for 30 minutes at 200 °C based on previously reported procedures. The active layer with a thickness of 100 nm was spin-coated from a chloroform solution of mixed DRCN5T:PC$_{70}$BM in a weight ratio of 15 mg ml$^{-1}$:12.5 mg ml$^{-1}$. After thermal annealing for 10 mins at 120 °C on a hotplate and subsequent solvent vapour annealing for 60 s in a Petri dish with saturated chloroform atmosphere, the substrates were transferred into a thermal evaporator (10$^{-7}$ mbar) for deposition of 10 nm MoO$_3$ and 80 nm Ag.

**Environmental Rig**

For all oxygen degradation studies, the OSCs were prepared the same way as mentioned above, stored in a nitrogen filled glovebox, and transferred to a sealed environmental box without exposure to ambient air. A constant flow of nitrogen was connected to the environmental box. The oxygen percentage was controlled by adjusting the relative flow rate of O$_2$ to N$_2$, and was continuously monitored by a zirconia sensor (Cambridge Sensotet, Rapidox 2100). All the samples were aged under simulated AM 1.5 sunlight at 100 mW cm$^{-2}$ irradiance (ABET Sunlite 11002 solar simulator).

**Current -voltage (J-V) measurements**

Current density–voltage (J-V) characteristics of the photovoltaic devices were measured using a Keithley 2450 Source Measure Unit. J-V curves of the photovoltaics were recorded under an AM 1.5 G light illumination (100 mW cm$^{-2}$) using an ABET Sunlite 11002 solar simulator, which was adjusted in intensity to compensate for the spectral mismatch between the spectral output and the spectral response of both the DRCN5T:PC$_{71}$BM device and the reference Si cell (NIST traceable, VLSI).

**External Quantum Efficiency (EQE)**

The EQE was measured with the monochromatic light of a halogen lamp from 350 to 800 nm, which was calibrated with a NIST-traceable Si diode (Thorlabs).

**UV-Vis absorption spectroscopy**

Films of DRCN5T and PC$_{70}$BM (15 mg mL$^{-1}$ and 12.5 mg mL$^{-1}$ respectively) were spin-coated on glass substrates in a N$_2$-filled glovebox and measured at room temperature with a JASCO UV-Vis V670 spectrometer.

**Transient absorption (pump-probe) spectroscopy**

Samples for TA measurements were prepared identically to the active layers of solar cells. Since this system does not exhibit electric-field induced charge generation[69] all measurements were performed on active layer films and reference films of the BHJ components. Seed pulses (800 nm, <100 fs) were generated at a repetition rate of 1 kHz by a Ti:Sapphire regenerative amplifier (Spectra-Physics Solstice, Newport Corporation), and routed towards an optical parametric amplifier (TOPAS, Light Conversion) coupled to a frequency mixer (NIRUVis, Light Conversion) to provide the 700 nm pump. The seed pulses were also routed towards a mechanical delay stage, then directed to a commercially available femtosecond transient absorption spectrometer (HELIOS, Ultrafast Systems). Therein, the pump (modulated at 500 Hz) and the broadband near-infrared (~800-1450 nm) probe were focused onto a ~0.5 mm$^2$ spot on the sample. Analyses were performed in Origin and MATLAB. The degradation of the films during the TA measurements was approximated by focussing a white light LED (Thorlabs MWWHD3) onto the measurement spot, where an intensity of 100 mW cm$^{-2}$ was assured by a silicon reference diode (NIST traceable, VLSI). The measurements were performed on unencapsulated films in the dry atmosphere of the optical lab.

**Photoemission spectroscopy**

Films of DRCN5T and PC$_{70}$BM (15 mg mL$^{-1}$ and 12.5 mg mL$^{-1}$, respectively) were spin-coated on glass/ITO/ZnO substrates in a N$_2$-filled glovebox. UPS measurements were performed using a double-differentially pumped He discharge lamp (hν = 21.22 eV) with a pass energy of 2 eV. During data collection the samples were biased at -5 V in order to measure the onset of the secondary electrons.

**Transient Photocurrent**

For transient photocurrent measurements, the light of an inorganic LED (Thorlabs TO-1 3/4, $\lambda$ = 465 nm) was pulsed by a function generator (Agilent/Keysight 33510B) and focused on the solar cell. An oscilloscope (Picoscope 5443A) with a 50 Ω terminator placed across the oscilloscope input was used to measure the transient photocurrent.

**X-Ray Diffraction (XRD)**

XRD measurements of the films on glass/ITO substrates were conducted on a Rigaku SmartLab diffractometer with a 9 kW rotating copper anode, equipped with a 0.5mmf collimator. The 2D diffraction pattern was collected with a HyPix3000 detector at a sample-detector distance of

110mm in a coupled Q/2Q scan from 0°-15°. Diffraction patterns were background corrected by substracting a measurement of the bare substrate.

**Acknowledgements**

We would like to kindly thank Prof. Uwe Bunz for providing access to the device fabrication facilities and Prof. Markus Motzkus for access to the ultra-fast laser lab. We thank the DFG (VA 991/2-1) for support. A.W. thanks Dr. Luis Perez-Lustres and Dr. Tiago Buckup for help with data analysis. This project has received funding from the European Research Council (ERC) under the European Union's Horizon 2020 research and innovation programme (ERC Grant Agreement n° 714067, ENERGYMAPS).